\documentstyle[11pt,newpasp,twoside,epsf]{article}

\begin{document}
\title{Studies of Mira and semiregular variables using visual databases}

\author{T.R. Bedding, B.C. Conn}
\affil{School of Physics, University of Sydney 2006, Australia}
\author{A.A. Zijlstra}

\affil{UMIST, Department of Physics, Manchester, UK}

\begin{abstract}
We use wavelets to investigate period and amplitude changes in Mira and
semiregular variables and found a variety of behaviours.  Period and
amplitude changes often go together, perhaps because changes in amplitude
are causing period changes via non-linear effects.
\end{abstract}

Wavelet analysis is useful for tracing changes in the periods and
amplitudes of long-period variables.  We have already published results for
the semiregular R~Dor, which appears to switch between two modes on a
timescale of decades (Bedding et al.\ 1998).  See also the work of
Szatm\'ary, Kiss and coworkers (Kiss et al.\ 1999 and these Proceedings),
and of Andronov (these Proceedings).

\begin{figure}
\plotone{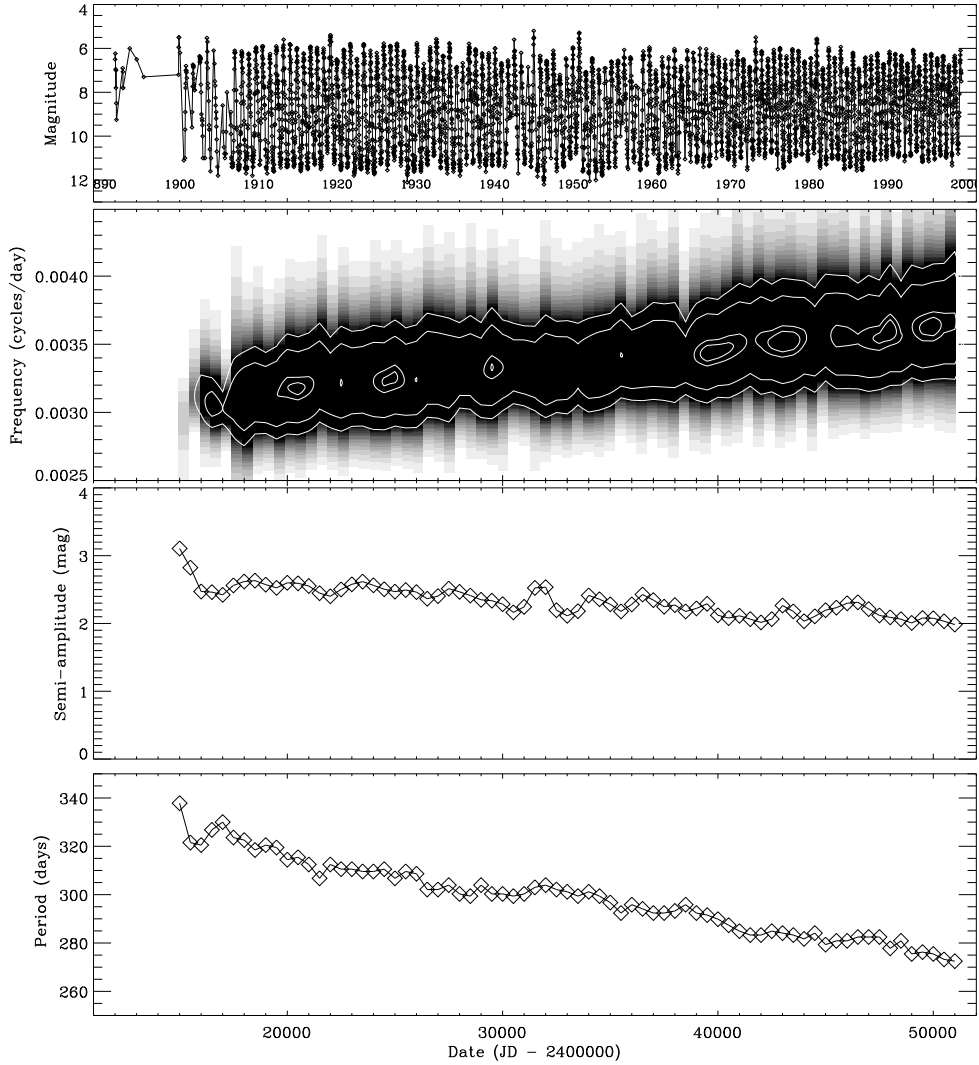}
\caption[]{\label{fig.raql}Analysis of visual observations of R~Aql.  The
top panel shows the light curve averaged in 5-day bins, from data supplied
by the AAVSO, AFOEV, VSOLJ and BAAVSS\@.  The second panel shows the WWZ
wavelet transform, while the third and fourth panels show time evolution of
the semi-amplitude and period, respectively.  As described by Bedding et
al.\ (1998), we use the wavelet implementation by Foster (1996), with the
semi-amplitude and period being derived from the WWA rather than the
WWZ\@.}
\end{figure}

\begin{figure}
\plotone{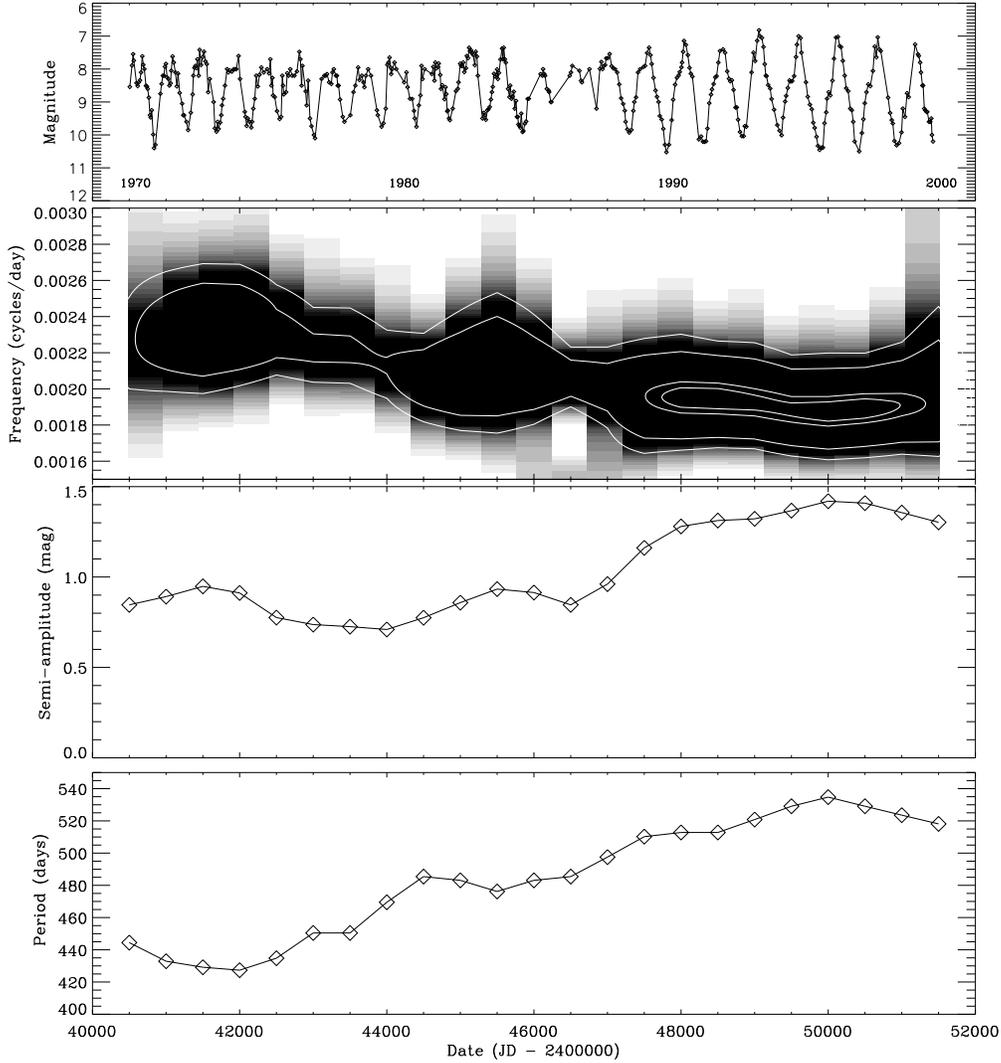}
\caption[]{\label{fig.bhcru}Same as Fig.~1 but for BH~Cru, from data
supplied by the RASNZ and averaged in 10-day bins.  Our analysis confirms
the suggestion by Bateson et al.\ (1988) that period of BH~Cru is
increasing.  The rate of period change ($\dot{P} = 0.012$\,d\,d$^{-1}$) is
the largest we have found for any Mira.  Meanwhile, as pointed out by
Walker et al.\ (1995) and confirmed by Hipparcos photometry, the double
maxima in the light curve have now disappeared.  As reviewed by Whitelock
(1998), the spectrum of this star has changed from O-rich to C-rich,
perhaps after dredge-up, and it is possible the period, amplitude and light
curve changes are also related to a recent dredge-up event.}
\end{figure}

In the case of Miras, a few stars are known to have long-term period trends
that may be related to adjustment after a helium-shell flash (Wood \& Zarro
1981).  A good example is R~Aql, which has a gradually decreasing period and,
as clear in Fig.~\ref{fig.raql}, a previously unnoticed decrease in
amplitude.  We find a similar effect in other Miras known to have changing
periods, such as W~Dra and BH~Cru (period and amplitude both increasing)
and R~Hya and T~UMi (periods and amplitudes decreasing).  The case of
BH~Cru is shown in Fig.~\ref{fig.bhcru}.

\begin{figure}
\plotone{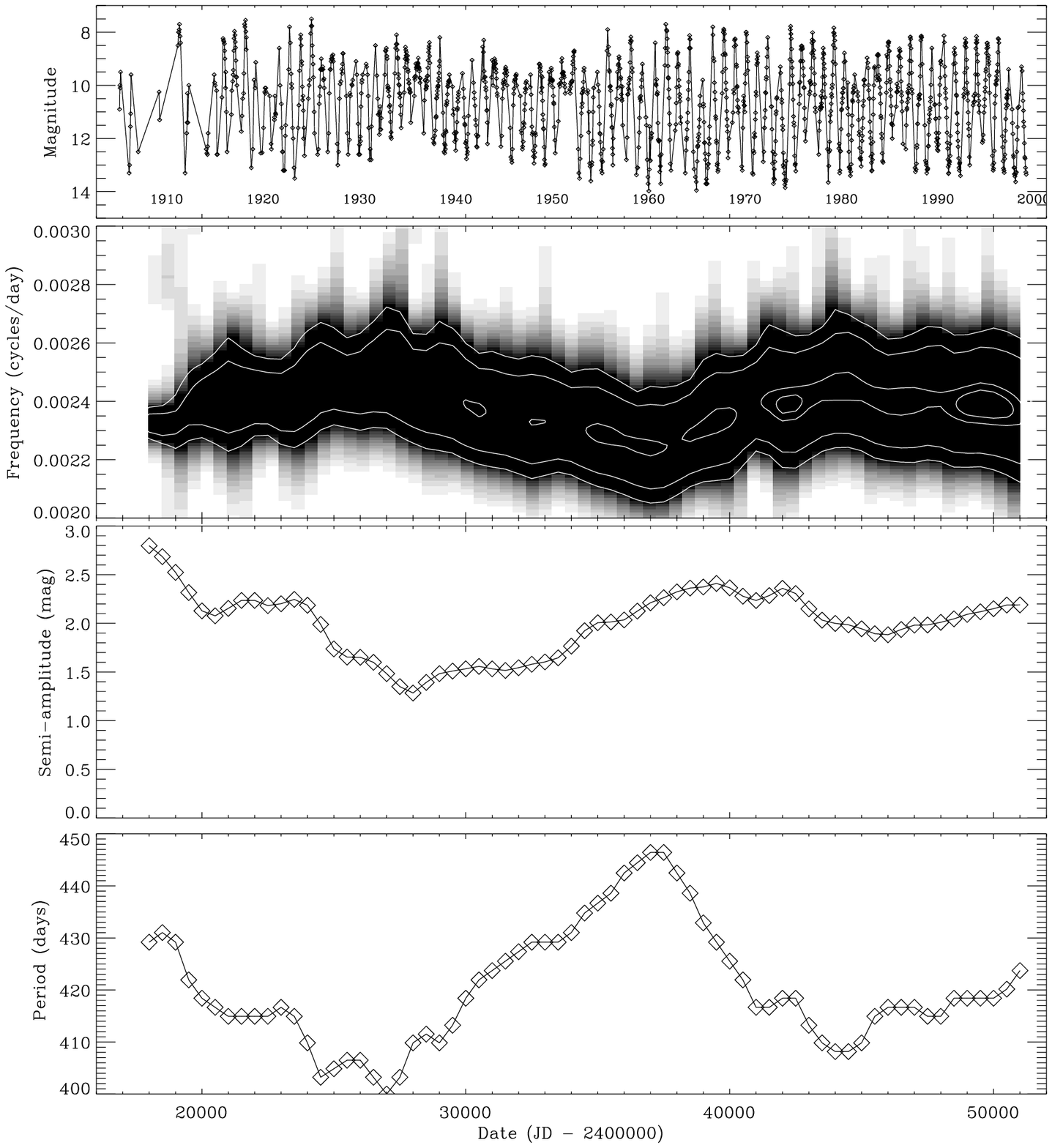}
\caption[]{\label{fig.sori}Same as Fig.~1 but for S~Ori, from data supplied
by the AAVSO, AFOEV and VSOLJ and averaged in 10-day bins.  Period changes
in this star are noted in the {\em General Catalogue of Variable Stars\/}
and also by Percy \& Au (1999).  Here, we see changes both in period and
period derivative ($\dot{P}$).  In fact, the period has not been constant
at any time in the last 100 years, with $\dot{P}$ changing sign every few
decades.  Note that the true behaviour is difficult to recognise in the
$O-C$ diagram (not shown), since a change in $\dot{P}$ must be deduced from
a change in curvature.  }
\end{figure}

Other stars such as S~Ori (Fig.~\ref{fig.sori}) show repeated period
changes, but still with matching amplitude changes.  We suggest that, at
least in some cases, the amplitude changes might {\em cause\/} the period
changes via non-linear effects.

\acknowledgments

We thank the many observers and those who maintain the visual databases of
the AAVSO, RASNZ, AFOEV, VSOLJ and BAAVSS\@.

%TRB thanks the Australian Research Council and the Science Foundation in
%the School of Physics for financial support.

\end{document}